\documentclass[preprint,12pt]{elsarticle}

\usepackage{amssymb}
\usepackage{graphicx}
\usepackage{amsmath}
\usepackage{amsthm}

\theoremstyle{plain}

\journal{...}

\begin{document}

\begin{frontmatter}



\title{Solitary wave solutions for  nonlinear 
partial differential equations containing monomials of odd and even grades
with respect to participating derivatives}


\author[1,2]{Nikolay K. Vitanov}
\author[3]{Zlatinka I. Dimitrova}

\address[1]{Institute of Mechanics, Bulgarian Academy of Sciences,
Akad. G. Bonchev Str., Bl. 4, 1113 Sofia, Bulgaria}
\address[2]{Max-Planck-Institute for the Physics of Complex Systems,
Noethnitzerstr. 38, 01187 Dresden, Germany}
\address[3]{"G. Nadjakov" Institute of Solid State Physics, Bulgarian Academy of 
Sciences, Blvd. Tzarigradsko Chaussee 72, 1784 Sofia, Bulgaria}

\begin{abstract}
We apply the method of simplest equation  for obtaining exact solitary 
traveling-wave solutions of  nonlinear partial differential equations that
contain monomials of odd and even grade with respect to participating 
derivatives. We consider first the general case of presence of monomials of
the both (odd and even) grades and then turn to the two particular cases of 
nonlinear equations that contain only monomials of odd grade or only monomials 
of even grade. The methodology is illustrated by numerous examples.
\end{abstract}


\end{frontmatter}



\section{Introduction}
Traveling waves exist in many natural systems. Because of this
traveling wave solutions of the nonlinear partial differential equations are
studied much in the last decades \cite{scott}-\cite{kudr90} and effective
methods for obtaining such solutions for integrable systems are
developed \cite{ablowitz2}, \cite{hirota}.  Our discussion below
will be based on a specific approach 
for obtaining exact special solutions of nonlinear PDEs: the method of simplest 
equation and its version called modified method of simplest
equation \cite{kudr05x} - \cite{vit11}. The method of simplest equation is 
based on  a procedure analogous to the first 
step of the test for the Painleve property \cite{kudr05}-\cite{k10}.                          
In the version of the method called  modified method of 
the simplest equation \cite{vdk10, vit11} this procedure is substituted by  
the concept for the balance equation. This version of the method of simplest 
equation has been successfully applied for obtaining exact traveling wave 
solutions of  numerous nonlinear PDEs such as versions of generalized 
Kuramoto - Sivashinsky equation, reaction - diffusion  equation, 
reaction - telegraph equation\cite{vdk10}, \cite{vd10}
generalized Swift - Hohenberg equation and generalized Rayleigh equation 
\cite{vit11}, extended Korteweg-de Vries equation \cite{v13}, etc. 
\cite{vit11b} - \cite{v13a}.
\par
The method of simplest equation works as follows. By means of an appropriate 
ansatz the solved   nonlinear PDE is reduced to a nonlinear ODE
\begin{equation}\label{i1}
P \left( u, u_{\xi},u_{\xi \xi},\dots \right) = 0.
\end{equation}
Then the finite-series solution 
$u(\xi) = \sum_{\mu=-\nu}^{\nu_1} p_{\mu} [f (\xi)]^{\mu}$
is substituted in (\ref{i1}). $p_\mu$ are coefficients and $f(\xi)$ is 
solution of simpler ordinary differential equation called simplest equation. 
Let the result of this substitution be a polynomial of $f(\xi)$. $u(\xi)$ is a 
solution of Eq.(\ref{i1}) if all coefficients of the obtained polynomial of 
$f(\xi)$ are equal to $0$. This condition leads to a system of nonlinear 
algebraic equations. Each non-trivial solution of this  system  corresponds 
to a solution of the studied  nonlinear PDE.
\par
Below we shall consider traveling-wave solutions 
$u(x,t)=u(\xi)=u(\alpha x + \beta t)$
constructed on the basis of the  simplest equation
\begin{equation}\label{a1}
f_{\xi}^2 = 4 (f^2 - f^3),
\end{equation}
which solution is $f(\xi)=\frac{1}{\cosh^2(\xi)}.$
$\alpha$ and $\beta$ are parameters. In addition we shall use the following
concept of grade of monomial with respect to participating derivatives. 
Let us consider polynomials that are linear combination of
monomials where each monomial  contains product of terms consisting of  
powers of derivatives of different orders. This product of terms can be 
multiplied by a polynomial of $u$. Let a term from a monomial contain $k$-th 
power of  a derivative of $l$-th order. \emph{We shall call the product $kl$ 
grade of the term with respect to participating derivatives. The sum of these 
grades of all terms of  a monomial will be called grade of the monomial with 
respect to participating derivatives.} The general case is
\textbf{(1):} The polynomial contains monomials that contain derivatives 
have odd or even grades with respect to participating derivatives. There are two particular cases:
\textbf{(1A):} All monomials that contain derivatives have odd grades with 
respect to participating derivatives; \textbf{(1B):}All monomials that contain 
derivatives have even grades with respect to participating derivatives.
We shall formulate our main result for the general case \textbf{(1)}
and then we shall demonstrate several applications for the cases \textbf{(1),
(1A), (1B)}.
\section{Main result}
Below we search for solitary wave solutions for the class of nonlinear PDEs that
contain  monomials of derivatives  which  order with respect to participating
derivatives is even and  monomials of derivatives  which  order with respect to 
participating derivatives is odd.
\par
\textbf{Theorem:}
\emph{
Let $\cal{P}$ be a polynomial of the function $u(x,t)$ and its derivatives.
$u(x,t)$  belongs to the 
differentiability class $C^k$, where $k$ is the highest order of derivative
participating in $\cal{P}$. $\cal{P}$ can contain some or all of the following 
parts:
\textbf{(A)}
polynomial of $u$;
\textbf{(B)}
monomials that contain derivatives of $u$ with respect to $x$ and/or products of
such derivatives. Each such monomial can be multiplied by a polynomial
of $u$;
\textbf{(C)}
monomials that contain derivatives of $u$ with respect to $t$ and/or products of
such derivatives. Each such monomial can be multiplied by a polynomial
of $u$;
\textbf{(D)}
monomials that contain mixed derivatives of $u$ with respect to $x$ and $t$ and/or products of such derivatives. Each such monomial can be multiplied by a polynomial
of $u$;
\textbf{(E)}
monomials that contain products of derivatives of $u$ with respect to $x$ 
and  derivatives of $u$ with respect to $t$. Each such monomial can be multiplied by a polynomial of $u$;
\textbf{(F)}
monomials that contain products of derivatives of $u$ with respect to $x$ 
and mixed derivatives of $u$ with respect to $x$ and $t$. 
Each such monomial can be multiplied by a polynomial of $u$;
\textbf{(G)}
monomials that contain products of derivatives of $u$ with respect to $t$ 
and mixed derivatives of $u$ with respect to $x$ and $t$. 
Each such monomial can be multiplied by a polynomial of $u$;
\textbf{(H)}
monomials that contain products of derivatives of $u$ with respect to 
$x$, derivatives of $u$ with respect to $t$ and mixed derivatives of 
$u$ with respect to $x$ and $t$. Each such monomial can be multiplied by 
a polynomial of $u$.}
\par
\emph{
Let us consider the nonlinear partial differential equation 
\begin{equation}\label{tb1}
{\cal{P}}=0.
\end{equation}
We search for solutions of this equation of the kind 
$u(\xi)=u(\alpha x + \beta t) = u(\xi) = \gamma f(\xi)$, 
where $\gamma$ is a parameter and $f(\xi)$ is solution of the simplest equation
$f_{\xi}^2 = 4 (f^2 - f^3)$. The substitution of this solution in Eq.(\ref{tb1})
leads to a relationship $\cal{R}$  of the kind
\begin{equation}\label{tb5}
{\cal{R}}= \sum_{i=0}^N C_i f(\xi)^i +f_{\xi} \left(\sum_{j=0}^M D_j f(\xi)^j \right)
\end{equation}
where $N$ and $M$ are natural numbers depending on the form of the polynomial $\cal{P}$ .
The coefficients $C_i$ and $D_j$ depend on the parameters of Eq.(\ref{tb1}) and on $\alpha, \beta$
and $\gamma$. Then each nontrivial solution of the nonlinear algebraic system
\begin{equation}\label{tb7}
C_i = 0, i=1,\dots,N; \ \ \ D_j = 0, j=1,\dots,M
\end{equation}
leads to solitary wave solution of the nonlinear partial differential equation (\ref{tb1}).
}
\begin{proof}
First we shall prove the following:\\
\emph{
Let $f(\xi)$ be a solution of the nonlinear ordinary differential equation
$f_{\xi}^2 = 4 (f^2 - f^3)$.
Then the even derivatives of $f(\xi)$ contain only a polynomial of $f(\xi)$.
The odd derivatives of $f(\xi)$ contain a polynomial of $f(\xi)$ multiplied
by $f_{\xi}$.}
\par
We shall use the method of induction.
The above statement is true for the second, third, and the fourth 
derivative of $f(\xi)$. Let the $2n$-th derivative of $f$ be a polynomial 
of $f$: $
\frac{d^{2n} f}{d \xi^{2n}} = F(f).
$.
Then
$
\frac{d^{2n+1} f}{d \xi^{2n+1}} = \frac{dF}{df} \frac{df}{d\xi},
$
and
$
\frac{d^{2n+2} f}{d \xi^{2n+2}} = \frac{d^2F}{df^2} \left( \frac{df}{d\xi} \right)^2 +
\frac{dF}{d\xi} \frac{d^2f}{d \xi^2} = F^*(f),
$
because of the fact that $\left(\dfrac{df}{d\xi} \right)^2$ and $\dfrac{d^2f}{d \xi^2}$ 
are polynomials of $f$ (this follows from the definition of $f$). 
Thus each even derivative of $f$ is a polynomial of $f$ 
and each odd derivative of $f$ is a polynomial of $f$ multiplied by $f_{\xi}$.
\par
Let  the nonlinear PDE ${\cal{P}}=0$ be reduced by the traveling-wave ansatz 
to the nonlinear ODE
${\cal{Q}}=0$ where each monomial has odd or even grades with respect to 
the participating derivatives. For the monomials with odd grades the 
chosen kind of solution reduces further the corresponding 
part of $\cal{Q}$ to relationship of the kind $\sum_{i=0}^N C_i f(\xi)^i$. 
For the monomials with even grades the chosen kind of solution reduces 
further the corresponding part of  $\cal{Q}$ to relationship of the kind $f_{\xi} \left(\sum_{j=0}^M D_j f(\xi)^j \right)$.
In such a way $\cal{Q}$ is reduced to a relationship of the kind ${\cal{R}}=\sum_{i=0}^N C_i f(\xi)^i +f_{\xi} \left(\sum_{j=0}^M D_j f(\xi)^j \right)$. Then each nontrivial
solution of the nonlinear algebraic system $C_i=0$; $D_j=0$ (if such solution exists) leads to a solitary traveling wave solution of the nonlinear PDE ${\cal{P}}=0$.
\end{proof}
\par
What follows are several examples. Let us consider the equation
\begin{equation}\label{ec1}
\nu u_{ttt} + \pi u u_{xx} + \sigma u_t^2 + (\delta + \mu u)u_x + \omega u^3 =0.
\end{equation}
The substitution of the studied kind of solution in Eq.(\ref{ec1}) leads to an 
equation that contains  monomials of odd and even grade with respect to 
derivatives of $u(\xi)$. The application of the above theorem leads to the 
following system of nonlinear algebraic equations:
\begin{eqnarray}\label{ec3}
-12 \beta^3  \nu + \alpha \gamma \mu = 0; \
4 \beta^3  \nu + \alpha \delta  =0; \nonumber \\
-6 \pi \alpha^2  - 4 \beta^2  \sigma + \gamma \omega =0; \
 \pi \alpha^2  +  \beta^2  \sigma = 0.
\end{eqnarray}
One solution of the system (\ref{ec3}) is
\begin{eqnarray}\label{ec4}
\alpha = - \frac{1}{2} \frac{(-\sigma \delta^2 \nu^2)^{3/4}}{\delta \nu^2 \pi^{3/4}};
\ \ 
\beta = \frac{1}{2} \frac{(- \sigma \delta^2 \nu^2)^{1/4}}{\nu \pi^{1/4}};
\nonumber \\
\gamma = - \frac{3 \delta}{\mu}; \ \ \omega =  \frac{1}{6} \frac{\sigma \mu (-\sigma \delta^2 \nu^2)^{1/2}}{\delta \nu^2 \pi^{1/2}},
\end{eqnarray}
and the corresponding solitary wave is
\begin{equation}\label{ec5}
u(x,t) = - \frac{3 \delta}{\mu \cosh^2 \left[- \dfrac{1}{2} \dfrac{(-\sigma \delta^2 \nu^2)^{3/4}}{\delta \nu^2 \pi^{3/4}} x + \dfrac{1}{2} \dfrac{(- \sigma \delta^2 \nu^2)^{1/4}}{\nu \pi^{1/4}} t \right]}.
\end{equation}
\par
As second example we consider the equation
\begin{equation}\label{ec6}
\pi u^2 u_{xxxx} + \mu u u_{xtt} + \nu u_{xx} u_t^2 + + \sigma u^2 u_{xt} + 
\delta u_x u_tt + (\epsilon u + \kappa u^2)u_t + \theta u^3 =0.
\end{equation}
The application of the theorem (here and in the examples below we let the 
calculations to the interested reader) leads to the solitary wave:
\begin{eqnarray}\label{ec10}
u(x,t) = -  \frac{ 3 \epsilon(\delta + 2 \mu)}{[2 \kappa (\delta + \mu)]
\cosh^2 \left[- \dfrac{5^{3/4}}{10} \dfrac{\epsilon^{1/2}(-\nu)^{1/4}}{\pi^{1/4} 
(\delta + \mu)^{1/2}} x + \dfrac{5^{1/4}}{2} \dfrac{\pi^{1/4} \epsilon^{1/2}}{(\delta 
+ \mu)^{1/2}(-\nu)^{1/4}} t  \right]}. \nonumber \\
\end{eqnarray}
\section{The two particular classes \textbf{(1A)} and \textbf{(1B)}}
For the class of nonlinear PDEs that contain monomials of derivatives which  
order with respect to participating derivatives is odd only (particular class
\textbf{(1A)}) the above theorem
is valid when there are no terms of kind $\textbf{(A)}$ and  $D_i=0$.
Famous example of nonlinear PDE of this class is the 
Korteweg-de Vries equation
\begin{equation}\label{e1}
u_t + K u u_x + u_{xxx} =0,
\end{equation}
where $K$ is a constant (in the original equation $K=6$). Eq.(\ref{e1})
contains three monomials and all of them have odd grade with respect to the 
participating derivatives. The application of above theorem to Eq.(\ref{e1})
leads to the famous solitary-wave solution of the Korteweg-deVries equation.
\par
Another nonlinear PDE that belongs to this class
of equations is the generalized Degasperis-Procesi equation
\begin{equation}\label{dp1}
u_t + c_0 u_x + d u_{xxx} - h^2 u_{xxt} = \frac{\partial}{\partial x} [c_1 u^2 +
c_2 u_x^2 + c_3 u u_{xx}].
\end{equation}
When $c_0 = c_2 = c_3 = h =0$, $d=1$ and $c_1 = - K/2$  Eq.(\ref{dp1}) is reduced to 
the Korteweg-de Vries equation. When $c_1 = -(3c_3)/(2h^2)$ and $c_2=c_3/2$
Eq.(\ref{dp1}) is reduced to the Camassa-Holm equation. Finally when $c_1 = -2c_3/h^2$
and $c_2=c_3$ Eq.(\ref{dp1}) is reduced to the Degasperis-Procesi equation 
which is used as model equation for propagation of 
shallow water waves over a flat bed \cite{const}.
The application of the theorem to Eq.(\ref{dp1}) leads to the solitary-wave 
solution
\begin{equation}\label{dp5}
u(x,t) = \frac{6 \alpha^2 (c_0 h^2+d)}{(4 \alpha^2 h^2-1) 
(c_1 - 2 \alpha^2 c_3)\cosh^2 \left[  \alpha x + \left( \dfrac{\alpha (4 \alpha^2 d + c_0)}{4 \alpha^2 h^2-1} t  \right) \right]}.
\end{equation}
As a last example we consider the nonlinear PDE
\begin{equation}\label{ex1}
\sigma_1 u_{xxxxx} + \sigma_2 u u_{xxx} + \sigma_3 u_{xxx} + \sigma_4 u_x u_{xx} + \sigma_5 u^2 u_x + \sigma_6 u u_x + u_t =0.
\end{equation}
The application of the theorem leads to the solitary wave solution
\begin{eqnarray}\label{ex6}
u(x,t) = \frac{12 \alpha^2 (20 \alpha^2 \sigma_1 + \sigma_3)}{(4 \alpha^2 \sigma_2 + 4 \alpha^2 \sigma_4 + \sigma_6)
\cosh^2 \left[ \alpha x + (-16 \alpha^5 \sigma_1-4 \alpha^3 \sigma_3)t \right]},
\end{eqnarray}
where
\begin{eqnarray}\label{ex4}
\alpha =  \frac{10^{1/2}}{20 \sigma_1 (10 \sigma_1 \sigma_5 - \sigma_2^2 - \sigma_2 \sigma_4)}
\Bigg[\sigma_1 (10 \sigma_1 \sigma_5 - \sigma_2^2 - \sigma_2 \sigma_4) \bigg(-20 \sigma_1 \nonumber \sigma_3 \sigma_5
\nonumber \\
- 5 \sigma_1 \sigma_4 \sigma_6 + 2 \sigma_2^2 \sigma_3 + 3 \sigma_2 \sigma_3 \sigma_4 + \sigma_3 \sigma_4^2 +
5 \sigma_1 \sigma_6 - \sigma_2 \sigma_3 -   \nonumber \\
\sigma_3 \sigma_4 \bigg( -40 \sigma_1 \sigma_5 + 4 \sigma_2^2 + 4 \sigma_2 \sigma_4 + \sigma_4^2\bigg)^{1/2} \bigg) \Bigg]^{1/2}, \nonumber \\
\end{eqnarray}
\par 
We would like to note the following.
Let us consider the nonlinear equations of the Korteweg-de Vries hierarchy of  
PDEs \cite{caud}
\begin{equation}\label{hier1}
K_x^{(n)} + u_t =0
\end{equation}
where $K^{(-1)} = -1$ and $K_x^{(n)} = - \dfrac{1}{4}[K_{xxx}^{(n-1)}) + 8 uK_x^{(n-1)}+4 u_x K^{(n-1)}]$.
All the equations from the KdV hierarchy   contain 
only monomials of odd grade with respect to the derivatives, i.e., are within the scope of the above theorem.
\par
For the second particular class \textbf{(1B)} of nonlinear PDEs the above 
theorem is valid with $C_i=0$. Famous representative of this class of 
equations is the Boussinesq equation 
\begin{equation}\label{beq1}
u_{tt} - u_{xx} - u_{xxxx} - K (u^2)_{xx} =0
\end{equation}
that is nonlinear model equation for shallow water waves valid
for weakly nonlinear long water waves. The application of the above theorem to 
Eq.(\ref{beq1}) leads to the well known solitary wave solution of the Boussinesq
equation.
\par
As another example we consider the nonlinear PDE 
\begin{equation}\label{ea1}
\mu u_{xt}+ \rho u_{tt} + \delta u u_{xx} + \epsilon u_{xxtt} + \pi u_x^2 +
\sigma u_t^2 =0.
\end{equation}
The application of the theorem leads to the following solitary wave solution
\begin{eqnarray}\label{ea5}
u(x,t) = -\frac{3 \rho}{2 \sigma} \left( \frac{5 [3 \delta \rho (4 \pi \sigma + 3 \delta \rho)]^{1/2}+ 15 \delta \rho + 4 \pi \sigma}{ [ 3 \delta \rho (4 \pi \sigma + 3 \delta \rho)]^{1/2} + 12 \delta \rho} \right) \times \nonumber \\
\dfrac{1}{ \cosh^{2} \left \{\dfrac{1}{2} \left(- \frac{\rho}{\epsilon} \right)^{1/2} x
+ \dfrac{1}{2\sigma (6 \epsilon)^{1/2}}  \left \{ - \sigma \bigg[\left[ 3 \delta \rho (4 \pi \sigma + 3 \delta \rho)\right ]^{1/2} + 3 \delta \rho \bigg]\right \}^{1/2} t \right \}}.
\nonumber \\
\end{eqnarray}
\section{Concluding remarks}
In this article we   formulate an approach for obtaining 
solitary wave solutions of a large  class of nonlinear PDEs.  
The discussed solitary  wave solutions have 3 parameters: $\alpha$, $\beta$, 
and $\gamma$. If the system of nonlinear algebraic equations contains 
3 equations then these 3 parameters can
be expressed by the parameters of the nonlinear PDE and there will be no need 
of relationships between the parameters of the nonlinear PDE. If the system of 
nonlinear algebraic equations contain more than 3 equations then there are also
relationships among the parameters of the solved nonlinear PDE. 
\par
Finally we note that the discussed methodology is a simple and effective
tool for obtaining exact solitary traveling wave solutions of numerous nonlinear
partial differential equations.
%
%






\end{document}